\documentclass[aps,pra,twocolumn,showpacs]{revtex4-1}
\usepackage{amsmath}
\usepackage{dsfont}
\usepackage{graphicx}
\newcommand{\bra}[1]{\langle {#1} |}
\newcommand{\ket}[1]{| {#1} \rangle}
\newcommand{\ah}{\hat{a}}
\newcommand{\Ph}{\hat{\Psi}}
\newcommand{\alh}{\hat{\alpha}}
\newcommand{\beh}{\hat{\beta}}
\newcommand{\rh}{\hat{b}}
\newcommand{\lh}{\hat{d}}
\newcommand{\eb}{\bar{\epsilon}}
\newcommand{\Hc}{{\hat{\cal H}}}
\newcommand{\id}{\mathds{1}}

\newcommand{\br}{{\bf r}}
\newcommand{\Phih}{\hat{\Phi}}

\makeatletter
\@ifundefined{definecolor}
 {\usepackage{color}}{}
\usepackage[colorlinks=true,linkcolor=blue,urlcolor=blue,citecolor=blue]{hyperref}

\begin{document}
\title{Edge-state instabilities of bosons in a topological band}
\date{\today}
\author{Ryan Barnett}
\affiliation{Department of Mathematics, Imperial College London,
London SW7 2AZ, United Kingdom}

\begin{abstract}
In this work, we consider the dynamics of bosons in bands with
non-trivial topological structure.  In particular, we focus on the
case where bosons are prepared in a higher-energy band and allowed to evolve.
The Bogoliubov theory about the initial state can have a dynamical
instability, and we show that it is possible to achieve the
interesting situation where the topological edge modes are unstable
while all bulk modes are stable.  Thus, after the initial preparation, 
the edge modes will become rapidly populated.  We illustrate this
with the Su-Schrieffer-Heeger model which can be realized
with a double-well optical lattice and is perhaps the simplest model
with topological edge states.  This work provides a direct physical
consequence of topological bands  whose properties are often not of
immediate relevance for the near-equilibrium properties of bosonic systems.
\end{abstract}
\pacs{67.85.-d, 03.75.-b, 37.10.Jk, 71.45.Lr}
\maketitle

\section{Introduction}
Bloch bands with non-trivial topological structure have been found to
have important physical consequences for a variety of fermionic
condensed matter systems including 1d conjugated polymers
\cite{heeger88}, quantum Hall systems \cite{thouless82}, and
topological insulators \cite{hasan10,qi11}.  Of central importance in
each of these systems is the presence of topologically protected edge
modes.  There currently are growing efforts to create and understand
bosons in non-trivial topological Bloch bands through the use of
ultracold atomic systems in optical lattices 
\cite{zhao11, hormozi12, cooper12, barnett12, yao12, price12, kjall12,
atala12,zhu13,ganeshan13, grusdt13, deng13}.  One of the promising
routes to such a realization is through the use of synthetic gauge
fields \cite{dalibard11, galitski13}, and very recently a physical
realization
of the Hofstadter model has been achieved  \cite{aidelsburger13,miyake13}.
However, the physical consequences of topologically non-trivial bands
is less direct for the near-equilibrium properties of bosons than it
is for fermions since bosons will generally populate the lowest energy
single-particle states, while higher-energy topological edge states
will be unoccupied.  Experimental probes for edge states in such
systems typically involve directly exciting bosons from the condensate
into these modes (see, for instance, \cite{stanescu09}).

In this work, we consider the evolution of a bosonic system (with
topological edge modes) initially prepared in a higher energy band
which has a dynamical instability.  Dynamical instabilities, which
give exponential growth of unstable modes in a conservative system,
have received considerable experimental and theoretical attention with
ultracold atoms (see, for instance, \cite{wu01, fallani04, sadler06,
cherng08, bucker11, hui12, ozawa13}).  We show that it is possible to
have the interesting situation where the edge modes are unstable while
all of the bulk modes are stable.  Therefore, after the system is
allowed to evolve the bosons will rapidly (exponentially fast) occupy
the edge modes.  Apart from being an experimental probe of edge
modes, the present work, more interestingly, proposes a new type of
non-equilibrium dynamics where bosons under a dynamical instability
rapidly populate these modes.  While a `holy grail' of current 
efforts with bosons in topological bands is to realize
strongly-interacting fractional quantum Hall phases, the dynamics
proposed here exists in the more common  weakly-interacting regime.

\section{Bogoliubov-SSH Hamiltonian}
Our work is
partially inspired by the recent `twin atomic beams' experiment 
\cite{bucker11}.  In this experiment, bosons are initially prepared in
the first excited transverse mode of a tube-shaped quasi
one-dimensional trap and allowed to evolve.  The system exhibits a
dynamical instability and beams peaked at opposite momenta
propagate longitudinally.  We propose a variation of this set-up where
the system is instead prepared in a quasi one-dimensional optical
lattice potential which has single-particle states with non-trivial
topology. For the non-interacting theory, we take the Su-Schrieffer-Heeger (SSH) 
Hamiltonian \cite{su79,heeger88}
\begin{align}
\label{Eq:SSH} \Hc_0 = -J\sum_n \left[(1+\eb (-1)^n) (\ah_n^\dagger
\ah_{n+1} + {\rm H.c.}) - 2 \ah_n^\dagger \ah_n \right]
\end{align}
which describes bosons hopping in a double-well 1d optical lattice
\cite{strabley06}.  In (\ref{Eq:SSH}), $\eb$ gives the magnitude of
staggering in the hopping where $0 \le \eb \le 1$, and we have included an overall
shift in the chemical potential so that the lowest single-particle
energy is zero.  We choose to work with the SSH Hamiltonian
(\ref{Eq:SSH}) since it is perhaps the simplest model which possesses
topological edge modes.  
However, the main results of this
work are expected to hold for any system with edge modes.
Measuring the Zak phase of the Rice-Mele model (which is the SSH model
with staggering in the onsite energies) was the focus of a recent
experimental work \cite{atala12}).  Methods of realizing 
fractionalized excitations of ultracold fermions in the SSH model were also proposed
in \cite{Ruostekoski02, Ruostekoski03}.

In \cite{schomerus13}, it was shown that adding an additional term
to (\ref{Eq:SSH}) with imaginary staggering in the on-site potential of the form
$\Hc'=i|\delta|\sum_n (-1)^n \ah_n^\dagger \ah_n$ can give exponential
growth of the edge modes.  Such a term will render the full
Hamiltonian non-hermitian, but can arise effectively for photons in
waveguides.  Here, we instead consider a fully hermitian Hamiltonian
$\Hc = \Hc_0 + \Hc_1$ that arises naturally when performing a
Bogoliubov expansion about the initial state of the
aforementioned experiments where
\begin{align}
\label{Eq:an}
\Hc_1=\sum_n \left[ (u -\Delta) \ah_n^\dagger \ah_n +
\frac{1}{2} u (\ah_n \ah_n + {\rm H.c.}) \right]
\end{align}
contains the anomalous terms.  In (\ref{Eq:an}), $u=U n_0$ 
where $U$ is the on-site Hubbard interaction and
$n_0$ is the average number of particles per site for the initial
state.  The parameter $\Delta$ is related to the mean-field energy
difference between the initial excited and ground state.
That is, when
$\Delta=0$, one recovers the Bogoliubov theory of bosons condensed in
the SSH lattice, and when $\Delta>0$, there can be a dynamical
instability.  A derivation of this Hamiltonian is presented in the
Appendix.
We also point out that
Hamiltonians of this form also  arise
in the context of quenched spinor condensates
\cite{lamacraft07,stamper-kurn13}, 
but here we will focus on scalar condensates
prepared in a higher-energy band for definiteness.

\section{General formalism for quadratic Bosonic systems}
In the following, we will briefly describe the general methods used to
compute the dynamical instabilities in finite systems.  Considering a
lattice with $N$ sites, one can write the full Hamiltonian as $\Hc =
\frac{1}{2} \Ph^\dagger H \Ph$ where $\Ph=(\ah_1,\ldots, \ah_N,
\ah_1^\dagger,\ldots, \ah_N^\dagger)^T$ and the Bogoliubov de Gennes
(BdG) Hamiltonian $H$ is a $2N \times 2N$ Hermitian matrix which can
be directly determined from (\ref{Eq:SSH}) and (\ref{Eq:an}).  It is
straightforward to see that the solution to the Heisenberg equations
of motion $i \hbar \partial_t \Ph = [ \Ph, \Hc]$ is given by
\begin{align}
\label{Eq:evolve} 
\Ph(t) = e^{-\frac{i}{\hbar} \tau_z H t} \Ph(0).
\end{align} 
Here, $\tau_\alpha=\sigma_\alpha \otimes \id$ where $\sigma_\alpha$
are Pauli matrices (where $\alpha$ can be $x$, $y$, or $z$) and $\id$
is the identity matrix.  Since $\tau_z H$ is in general not Hermitian,
it may have complex eigenvalues.  When this occurs, the system is said
to have a dynamical instability.  This can be contrasted with the
analogous problem of quadratic fermionic Hamiltonians which cannot
have complex modes and therefore will never have a dynamical
instability.  To further understand (\ref{Eq:evolve}), we consider
the BdG equation
\begin{align}
\label{Eq:BdG}
\tau_z H \psi_{i\pm} = \pm E_i \psi_{i\pm}
\end{align}
where $\psi_{i\pm} $  is a  $2N$ dimensional  eigenvector.

For our problem, as is verified numerically, the
eigenvalues of $\tau_z H$ are either purely real or imaginary.  We
will consider these cases separately.  Because of the symmetry
$
\tau_x H \tau_x= H^*, 
$
the real eigenvalues occur in pairs $\pm E_i$ (as already indicated in
(\ref{Eq:BdG})) with $\psi_{i-} = \tau_x \psi_{i+}^*$.  
For the real case, eigenvectors can be normalized as
$\psi_{i+}^\dagger \tau_z \psi_{i'+} = \delta_{ii'}$,
$\psi_{i-}^\dagger \tau_z \psi_{i'-} = -\delta_{ii'}$, and we also
have that $\psi_{i+}^\dagger \tau_z \psi_{i'-} =0$ \cite{blaizot86}.
We now introduce the operators $\alh_i = \psi_{i+}^\dagger \tau_z\Ph$
(or $\alh_i^\dagger=-\psi_{i,-}^\dagger \tau_z \Ph$) which can be seen
to satisfy bosonic commutation relations and diagonalize the stable
portion of the full Hamiltonian.

We now consider the imaginary eigenvalues of $\tau_z H$.  These will
also occur in $\pm$ pairs since for a right eigenvector $\psi_{i+}$ of
energy $E_i$, we can obtain a left eigenvector
$\psi_{i+}^\dagger\tau_z$ of energy $E_i^*=-E_i$.
These eigenvectors  can be
normalized as $\psi_{i+}^\dagger \tau_z \psi_{i'-} = i
\delta_{ii'}$ and we also have that $\psi_{i+}^\dagger\tau_z
\psi_{i'+} =\psi_{i-}^\dagger\tau_z \psi_{i'-}=0$ \cite{blaizot86}.
The operators defined as $\hat{x}_i= i \psi_{i-}^\dagger \tau_z \Ph$
and $\hat{p}_i = -i \psi_{i+}^\dagger \tau_z \Ph$ can then be seen to
satisfy canonical commutation relations $[\hat{x}_i, \hat{p}_{i'}] = i
\delta_{ii'}$.  Equations can be simplified further by introducing the
bosonic operators $\beh_i=\frac{1}{\sqrt{2}}(e^{-i\frac{\pi}{4}}
\hat{x}_i +e^{i\frac{\pi}{4}}\hat{p}_i)$.  Then, together with the
results from the real eigenvalues, we are able to rewrite the full
Hamiltonian in the quasi-diagonal form
\begin{align}
\label{Eq:canonical}
\Hc= \sum_i  (E_i+1/2) \alh_i^\dagger \alh_i + \sum_i' \frac{1}{2} |E_i|
(\beh_i \beh_i + \beh_i^\dagger \beh_i^\dagger )
\end{align}
where the first summation is over stable modes and the second
(primed) summation is over unstable modes
\footnote{This can be verified by inserting the resolution of the
  identity
$
\id = \sum_i (\psi_{i+} \psi_{i+}^\dagger - \psi_{i-} \psi_{i-}^\dagger) \tau_z + 
\sum_i' (i \psi_{i+} \psi_{i-}^\dagger - i \psi_{i-} \psi_{i+}^\dagger)  \tau_z
$
after the propagator in (\ref{Eq:evolve}), rewriting $\Ph(t)$ in terms
of $\alh_i, \beh_i$, and comparing with the Heisenberg equations of
motion from (\ref{Eq:canonical}).}. 
Note that a Hamiltonian that has a dynamical instability cannot be
brought fully to diagonal form.  It is straightforward to adapt the
above formalism to the momentum basis or continuous systems, which
will be done in the following paragraphs.

\section{Bulk modes}
\begin{figure}
\includegraphics[width=3in]{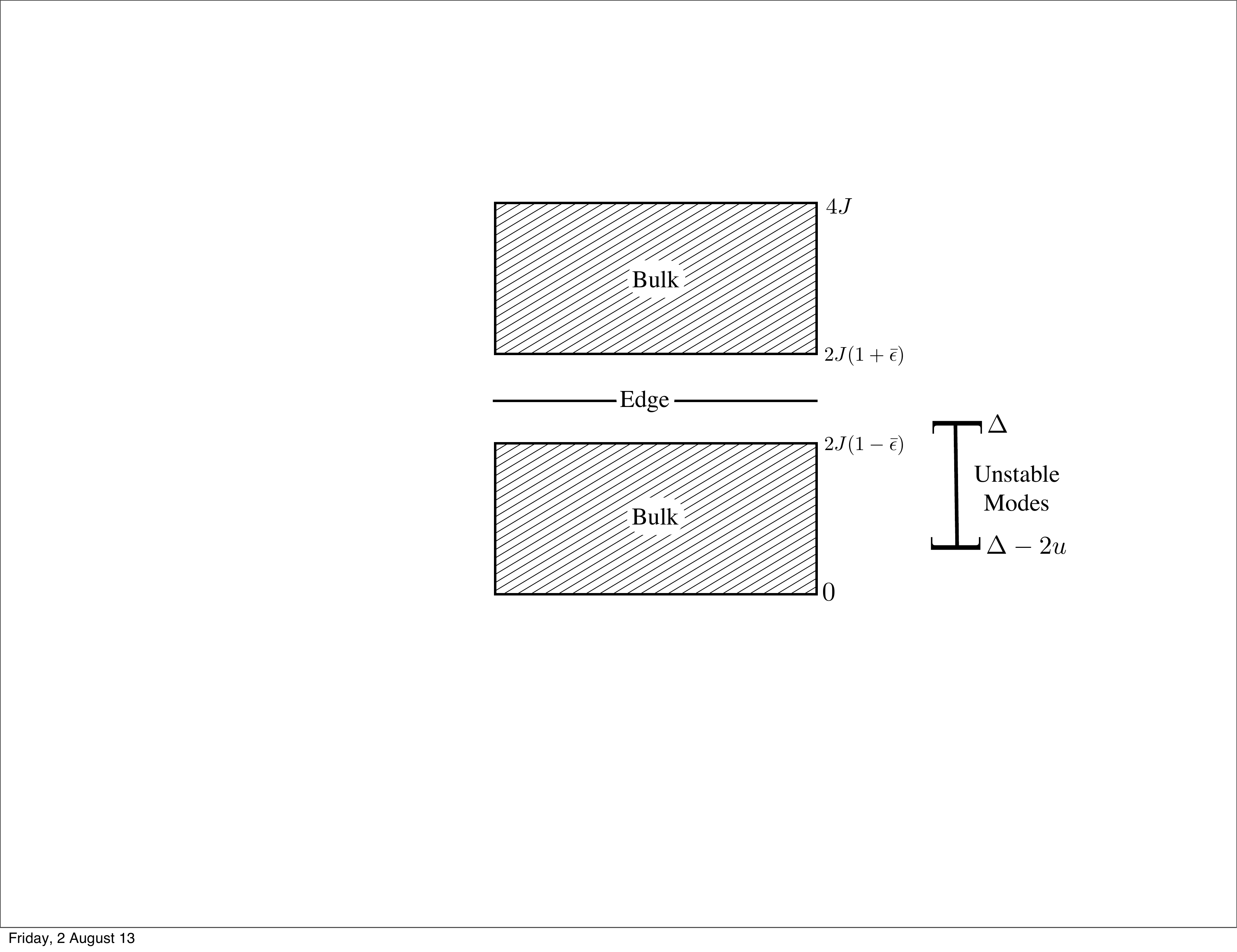}
\caption{Schematic description of dynamical instability.  Left:
  energies of the \emph{non-interacting} bulk and edge states.  Right: region
where dynamical instability occurs for a particular value of
parameters.  For the parameters illustrated in this figure, only the
bulk modes in the lower band will be unstable.}
\label{Fig:1}
\end{figure}

We will now focus on performing the above explicitly for our model.
The bulk modes are expected to be well-approximated by applying
periodic boundary conditions.  It is easiest to begin by working in a
basis which diagonalizes the non-interacting SSH Hamiltonian
(\ref{Eq:SSH}).  It is straightforward to find that the
non-interacting band energies are
\begin{align}
\label{Eq:bands} 
\xi_k^{(1,2)} = 2J \mp 2J \sqrt{\cos^2(k) + \eb^2
\sin^2(k)}
\end{align} where $k$ is restricted to the reduced Brillouin zone
$[-\pi/2,\pi/2]$ (for simplicity we set the lattice constant to
unity).  These non-interacting bulk bands (\ref{Eq:bands}) satisfy
$\xi_k^{(1)} \in [ 0 , 2J(1-\eb)]$, $\xi_k^{(2)} \in [ 2J(1+\eb),
4t]$.  Writing (\ref{Eq:SSH},\ref{Eq:an}) in this basis, one finds
that the
Bogoliubov energies are the eigenvalues of $\tau_z H_k$ where
$H_k=(2J+u-\Delta) \id \otimes \id +\frac{1}{2}\left(\xi_k^{(2)} -
\xi_k^{(1)}\right) \id \otimes \sigma_z + u \sigma_x \otimes \id$
($\id$ here is the $2\times2$ identity matrix).  Through this,  
the bulk energies are evaluated to be
\begin{equation}
\label{Eq:bulken}
E_k^{(n)} =  \sqrt{ (\xi_k^{(n)} - \Delta)(\xi_k^{(n)} - \Delta + 2u)}.
\end{equation}
When $\Delta=0$, (\ref{Eq:bulken}) is real for all values of $k$ and one
recovers the phonon modes of bosons condensed in the SSH lattice
potential which has sound speed $c=\sqrt{2uJ(1-\eb^2)}/\hbar$.
On the other hand, when $\Delta>0$, it is possible to have imaginary
values of (\ref{Eq:bulken}).  In particular, a bulk mode that satisfies
\begin{equation} 
\label{Eq:bulkcond}
\Delta - 2u < \xi_k^{(n)} < \Delta,
\end{equation} 
will correspond to a dynamical instability.  
The condition for these modes to be stable/unstable is shown
pictorially in Fig.~\ref{Fig:1}.

\section{Edge modes}
We now move on to a description of the edge modes of our system.  It
is well known that the non-interacting SSH model can have topological
mid-gap states localized at the edge of the system \cite{heeger88}.
The existence of such states are predicted by the non-trivial Zak
phase of the bulk banks of the SSH model \cite{zak89}.
More precisely, an edge mode of energy $2J$ will be present if a site
at the boundary is connected with the smaller of the two hopping
parameters, $J(1-|\eb|)$ and $J(1+|\eb|)$.  We will restrict our
attention to systems with an odd number of sites $N$, with the first
site labelled by $n=1$, thus ensuring precisely one edge mode in the
non-interacting spectrum at the left-hand side of the system.

We now proceed to analyze the fate of the edge mode when the anomalous
portion of the Hamiltonian (\ref{Eq:an}) is accounted for.  As a model
for the edge mode, we allow $\eb$ in (\ref{Eq:SSH}) to depend on
position and thus replace $\eb \rightarrow \epsilon_n$.  We take
$\epsilon_n$ to have a localized ``kink'' so that $\epsilon_n
\rightarrow \pm \bar{\epsilon}$ far to the right (left) of the 
kink.
That is, moving across the kink changes the sign of $\epsilon_n$.
This lattice defect will bind a state which is topologically
equivalent to an edge mode \cite{jackiw76, yao09, hasan10}.  
We next decompose $\ah_n$ as
\begin{align}
\label{Eq:andreev}
\ah_n = e^{i \frac{\pi}{2} n} \rh_n+ e^{-i \frac{\pi}{2} n} \lh_n
\end{align}
where the right and left movers, $\rh_n$ and $\lh_n$, are taken to be
slowly varying on the scale of the lattice constant (we will drop
second second and higher order derivatives in the continuum limit of these terms).  This
approximation is expected to be valid for small $\eb$ since for this
case, the edge mode is primarily composed of states at the band edges \cite{heeger88}.  Inserting
(\ref{Eq:andreev}) into (\ref{Eq:SSH},\ref{Eq:an}), taking the
continuum limit $\rh_n, \lh_n \rightarrow \rh(x), \lh(x) $, and
dropping higher  derivatives, we find $\Hc =\frac{1}{2} \int dx
\Ph^\dagger (x) H(x) \Ph(x) $ where $\Ph(x) = (\rh(x), \lh(x),
\rh^\dagger(x), \lh^\dagger(x))^T$ and here the BdG Hamiltonian is
\begin{align}
H=& (2J+u-\Delta) \id \otimes \id + 2J (-i \partial_x) \, \sigma_z \otimes
\sigma_z \\ &- 2 J \epsilon(x)  \,  \sigma_z \otimes \sigma_y + u \, \sigma_x
\otimes \sigma_x.
\notag
\end{align}
For simplicity, we choose the kink to be centered at $x=0$ and
therefore take $\epsilon(x)$ to be antisymmetric about this point.
Since $H$ only involves first order derivatives, the BdG equation for
positive energy, $\tau_z H(x) \psi_+(x)= E \psi_+(x)$, has the general
solution
\begin{align}
\label{Eq:wf}
\psi_+(x) = e^{\int_0^x dx' F(x')} \psi(0)
\end{align}
where
$
\allowbreak
F(x)=\frac{1}{2J}[i E \id \otimes \sigma_ z -i
  (2J+u-\Delta) \sigma_z \otimes \sigma_z +2J\epsilon(x) \id \otimes
  \sigma_x + iu \sigma_y \otimes \sigma_y ].
$

We search for a solution (\ref{Eq:wf}) which exponentially decays away from the kink
and thus require
\begin{align}
\label{Eq:b1}
F|_{\epsilon=\bar{\epsilon}} \; \psi_+(0) &=
  -\kappa_1 \psi_+(0) \\ 
\label{Eq:b2}
F|_{\epsilon=-\bar{\epsilon}} \; \psi_+(0) &=
  \kappa_2 \psi_+(0)
\end{align}
for $\kappa_1, \kappa_2>0$.
Subtracting (\ref{Eq:b1}) from (\ref{Eq:b2}) gives
\begin{align}
2\bar{\epsilon} \; \id \otimes \sigma_x \; \psi_+(0) = -(\kappa_1 + \kappa_2) \psi_+(0).
\end{align}
Therefore, $\psi_+(0) \propto \chi \otimes \chi_{-x}$ where
$\sigma_x \chi_{-x}=- \chi_{-x}$ and $\chi$ is to be determined.
Adding (\ref{Eq:b1}) and (\ref{Eq:b2}) with this condition on $\psi_+(0)$  then gives
\begin{align}
\notag
i  E \chi \otimes \chi_{x} - i   (2J+u-\Delta)& \sigma_z \chi \otimes
  \chi_x  - u \; \sigma_y \chi \otimes \chi_x  \\=& J (\kappa_2 -
\kappa_1) \chi \otimes \chi_{-x}
\label{Eq:add}
\end{align}
where $\sigma_x\chi_x=\chi_x$.  This forces
$\kappa_1=\kappa_2=\bar{\epsilon}$ and (\ref{Eq:add}) is simplified to
\begin{align}
\label{Eq:add2}
[(2J+u-\Delta)\sigma_z - iu \sigma_y]\chi=  E\chi.
\end{align}
From this, the energy of the edge mode immediately follows:
\begin{equation}
\label{Eq:edgeen}
E_{\rm edge} = \sqrt{(2J-\Delta)(2J-\Delta+2u)}
\end{equation}
which will be imaginary when
\begin{align}
\label{Eq:edgecond}
\Delta - 2u < 2J < \Delta.
\end{align}

By comparing (\ref{Eq:bulkcond}) and (\ref{Eq:edgecond}) one sees that
for certain parameters, it is possible to have the situation where the
edge mode is unstable but all bulk modes are stable.  The optimal
value of $\Delta$ for this to occur is $\Delta=\Delta_*\equiv 2J+u$ so that the
region of unstable modes is centered in the gap
(cf.~Fig.~\ref{Fig:1}).  Then it is clear that all of the bulk modes
will be stable if $u<2J \eb$.  At the optimal value of $\Delta$, we
also have $E_{ \rm edge} =
iu$ and the BdG wave function (\ref{Eq:wf}) can be found from
(\ref{Eq:b1},\ref{Eq:b2}) and takes on the relatively simple form
\begin{align}
\label{Eq:bdgwf}
\psi_+(x) ={\cal N} e^{-\bar{\epsilon} |x|}
  (\omega, \omega^{-3}, \omega^{-1},\omega^{3})^T
\end{align}
where $\omega=e^{i\frac{\pi}{4}}$ and ${\cal N}$ is a (real)
normalization constant.  The overall phase is chosen so that
$\psi_+^*(x)=\tau_x \psi_+(x)$.   A very similar analysis can be used
to find $\psi_{-}(x)$.

We now consider the experimentally relevant case of starting with a
vacuum state of $\ah_n$ bosons.  This occurs for the case when all
atoms are prepared in the higher-energy band \cite{bucker11} (see also
the Appendix).  Quantum 
fluctuations will trigger the evolution of this state (which at the
classical level is stationary) into the lower band.  To elucidate this
behavior, we consider the time dependence of the population per site
in the lowest band given by
\begin{align}
\label{Eq:gain1}
G_n(t) \equiv
\langle \ah_n^\dagger(t) \ah_n(t) \rangle 
\end{align}
where $\ah_n(t) = e^{\frac{i}{\hbar} \Hc t} \ah_n e^{-\frac{i}{\hbar} \Hc t} $ in the Heisenberg
picture and the expectation value is evaluated with the vacuum state
corresponding to zero initial population in the lower band.  
We take a finite system with an odd number of sites, so that the edge
mode decays to the right from site $n=1$ (note that we number the
lattice sites so that $n>0$).
We consider the case where only the edge
mode is unstable and take $\Delta=\Delta_*$, with $u<2J\eb$.  The BdG
wave function (\ref{Eq:bdgwf}) can then be used to find an expression
for the time-dependent population of the atoms in the lowest band at a
particular site when $|\bar{\epsilon}|\ll 1$. Reverting back to the
case of a discrete lattice, we find
\begin{align}
\label{Eq:gain2}
G_n(t) = 4 \eb
e^{-2\eb n} \sin^2\left(\frac{\pi n}{2}  \right) \sinh^2\left(\frac{u t}{\hbar}\right).
\end{align}
In deriving (\ref{Eq:gain2}) we have neglected the contribution from
the stable bulk modes which have oscillatory time dependence, and
whose relative contribution to (\ref{Eq:gain1}) becomes small for
$|E_{\rm edge}| t / \hbar \gg 1$.  Note that the initial
vacuum state will generically have non-zero overlap with the bulk
states of system.  For these parameters, the bulk band with energy
$E_k^{(1)}$ will be energetically unstable, but the corresponding modes
will not grow since the total energy is conserved.
Eq. (\ref{Eq:gain2}) gives exponential
growth of the edge mode.  One should note, however, that when the
number of depleted bosons $\sum_n G_n(t)$ is on the order of the total
particle number, the Bogoliubov theory breaks down, and
(\ref{Eq:gain2}) is inapplicable.

\begin{figure}
\includegraphics[width=3.4in]{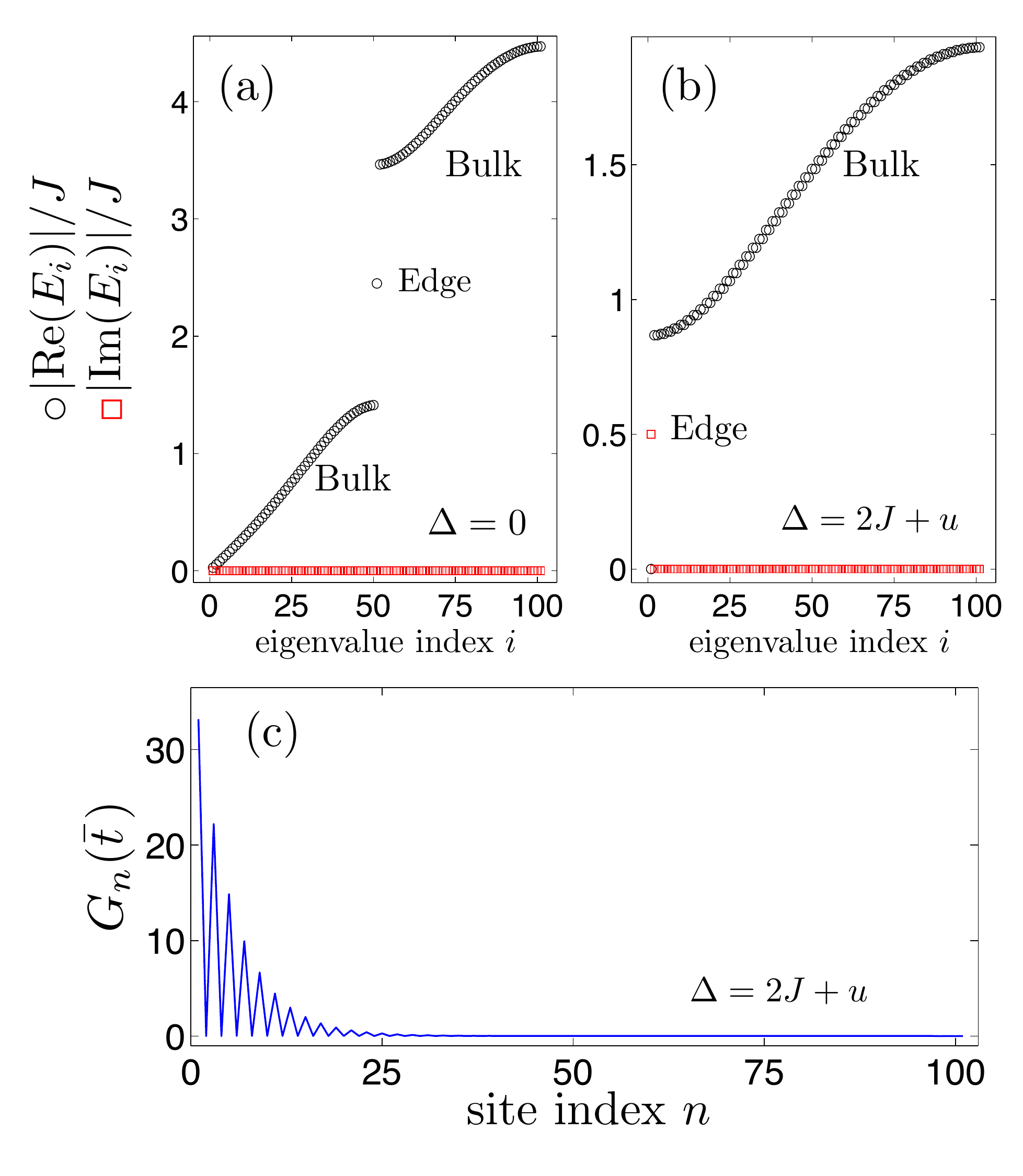}
\caption{(Color online) Top: the eigenvalues of $\tau_z H$ from direct
  diagonalization
of (\ref{Eq:SSH},\ref{Eq:an}) in ascending order 
for a lattice with $N=101$ sites and parameters $u/J=\eb=1/2$.  (a)
has $\Delta=0$ and so all modes are real while (b) has $\Delta=2J+u$.
(c):  The population per site after time $\bar{t}=30\hbar/J$ for  $\Delta=2J+u$.  So that
$G_n(\bar{t})$ extends over several lattice sites, the values $u/J=\eb=1/10$
were chosen in (c).}
\label{Fig:2}
\end{figure}

\section{Numerical diagonalization.}
We now move on to discuss the direct numerical diagonalization of
the $2N \times 2N$ BdG equation.
This will allow us to validate the analytic results found previously,
and also to access the regime where $\eb$ is not small.  Interestingly,
the conditions established previously for the edge state to by
stable/unstable remain accurate when $\eb$ is not small.
Results are shown in
Fig. \ref{Fig:2}.    Panel (a) shows the eigenvalues of the BdG equation for
the case of $\Delta=0$ which has all real eigenvalues, as expected.
These are the Bogoliubov energies of bosons
condensed in the ground state of the SSH model.  
In panel (b), the optimal value of $\Delta=\Delta_*$ is
chosen.  As is indicated by (\ref{Eq:bulkcond}), (\ref{Eq:edgecond}) and
confirmed by the diagonalization, the only imaginary eigenvalue is
associated with the edge mode.
The analytical expressions for the bulk and edge energies
(\ref{Eq:bulken},\ref{Eq:gain2}) show excellent agreement with the
results from the numerical diagonalization.
The population per site, $G_n(t)$, can be computed numerically from
(\ref{Eq:evolve}).  In panel (c), $G_n(t)$ is plotted, again for the
optimal value of $\Delta$.  For this panel, we set
$\bar{\epsilon}=1/10$ so that the edge mode extends across several
lattice sites, but is still well-localized about the LHS of the system 
($N=101 > 1/2\bar{\epsilon}$).
For the time shown in the plot, the
unstable edge has the dominant contribution to $G_n(t)$, and the
expression (\ref{Eq:gain2}) shows excellent agreement with the exact result
shown in the figure.

\section{Discussion and conclusion}  
The microscopic parameters entering (\ref{Eq:SSH}, \ref{Eq:an}) for
the proposed experiment of having an unstable edge mode but stable
bulk modes are within current experimental range.  The value of
$\Delta$ can be tuned over a wide range of values by changing the trap
confinement in the tight direction as discussed in the Appendix.
Though the experiment in \cite{bucker11} is done without an optical
lattice, their value of $\Delta$ is about a factor of four larger than
the mean field interaction energy.  Now consider tuning $\Delta$ to
its optimal value, $\Delta=\Delta_*=2J+u$, so that the region of
unstable modes as pictured in Fig.~\ref{Fig:1} is centered in the gap.
For non-zero $u$, the edge mode will have a dynamical instability.
For all of the bulk modes to be (dynamically) stable, we have the
further requirement $u<2J\bar{\epsilon}$.  This requirement is
consistent with the initial state being in the superfluid regime (away
from the Mott Insulator transition) and will occur when
$\bar{\epsilon}$ is not too small.

Our treatment of finite
systems with open boundary conditions is somewhat over-idealized in
that experiments in ultracold gases normally also involve a confining
potential $V(x_n)$ which adds the additional contribution $\Hc_{\rm
trap} = \sum_n V(x_n) \ah^\dagger_n\ah_n$ to the full Hamiltonian.
While a thorough treatment of the confining potential is beyond the
scope of the present work, we note that it is shown in
\cite{buchhold12} that many of the features of edge modes remain when
a harmonic confining potential is applied.  Alternatively, it is
possible to engineer sharp boundary conditions as described in
\cite{goldman10} which are very similar to open boundary conditions.
Finally, it may be possible to directly engineer a kink or domain wall
using the recently developed methods to address single sites of an
optical lattice \cite{bakr09, weitenberg11}.

In summary, we have described a method through which the topological
edge modes of a system can tuned to have a dynamical instability while
all the bulk modes remain stable.  This is particularly interesting
since  topological bands typically play an inessential role
in condensed bosonic systems.  To illustrate, motivated by its
simplicity, we have considered the SSH model with anomalous terms.  An
interesting avenue of future study will be to consider similar
preparations of two-dimensional Chern insulators \cite{haldane88}.  A
crucial difference with such systems is that the edge modes in the 2d systems have
dispersion and so only a portion of them will be unstable.  This work
only concentrated on the quadratic theory expanded about the initial
state, which will inevitably break down at sufficiently long times
when the number of depleted bosons becomes comparable to the number of
condensed bosons.  Another interesting future direction is to use more
elaborate theoretical techniques to explore such dynamics for longer times.

\acknowledgements
Funding from Imperial College London is gratefully acknowledged as
well as  the Aspen Center for Physics under Grant No. PHYS-1066293 
where part of this work was completed.
I would like to thank
Austen Lamacraft, 
Derek Lee,
 J.V.~Porto, and
Ari Turner for particularly helpful discussions.

\appendix

\section{Derivation of the Bogoliubov-SSH Hamiltonian}
In the following, we discuss a context in which the effective
Hamiltonian given in Eqns.~(1,2)  of the manuscript 
arises.  We consider a setup akin
to that used in the experiment in \cite{bucker11}, but in the presence
of a double-well optical lattice.  The full Hamiltonian of the system
is
\begin{align}
\label{Eq:H}
\notag
\Hc = \int d^3 r  & \left[ \Phih^\dagger(\br) \left(
  -\frac{\hbar^2}{2m}\nabla^2 +V_{\rm trap}(y,z) +V_{\rm lat}(x)  
\right.  \right. \\  & 
\left.  \left.  -\mu  \vphantom{\frac{\hbar^2}{2m}}\right) \Phih(\br) +\frac{g}{2}  \Phih^\dagger(\br) \Phih^\dagger(\br)  \Phih(\br)  \Phih(\br) \right].
\end{align}
In this equation, $V_{\rm trap}(y,z)=\frac{1}{2} m(\omega_y^2 y^2 +
\omega_z^2 z^2)$ is a harmonic trapping potential,
$V_{\rm lat}(x)$ is the optical lattice potential, $\mu$ is the
chemical potential, $g$ is the interaction parameter related to
the three-dimensional $s$-wave scattering length, and $\Phih(\br)$ are
the bosonic field operators:
$[\Phih(\br),\Phih^\dagger(\br')]=\delta(\br-\br')$.  For $V_{\rm
  lat}(x)$, 
we take a one-dimensional double-well lattice having, for
instance, the form
\begin{align}
V_{\rm lat}(x) = V_1\cos(Qx)+V_2 \cos(2Qx)
\end{align}
with $V_1,V_2>0$.
As in
\cite{bucker11}, we take tight confinement in the $y$ and $z$
directions with  $\omega_y \lesssim \omega_z$ and consider an
initial state that is in the first excited state of the trapping
potential.  We accordingly expand  the bosonic field operator  into the ground and
excited bands (where the excited band corresponds to the first excited
spatial mode of the trap): $\Phih(\br)=\Phih_g(\br) +\Phih_e(\br)$ where
\begin{align}
\Phih_g(\br)&=  \sum_n \phi_{g}(y,z)  
 w_{n} (x) \ah_{g,n} \\
\Phih_e(\br)&=\sum_n \phi_{e}(y,z) w_n(x)  \ah_{e,n}.
\end{align}
In this, the one-dimensional sum is over the lattice sites situated at
the minima of $V_{\rm lat}(x)$ in
the $x$-direction.  The Wannier orbital $w_{n}(x)$ is centered at
site $n$,  
and  $\phi_{g}(y,z)$ and $\phi_{e}(y,z)$ are ground and
first excited spatial modes of the trap which are well approximated by harmonic
oscillator wave functions for strong confinement.    The orbitals
$\phi_{g}(y,z)$ and $\phi_{e}(y,z)$ are taken to be real and are
symmetrical under $z \rightarrow -z$, while
$\phi_{g}(y,z)=\phi_{g}(-y,z)$ and $\phi_{e}(-y,z)=-\phi_{e}(y,z)$.
Inserting
$\Phih(\br)=\Phih_g(\br) +\Phih_e(\br)$ 
into $\Hc$, using the parity of $\phi_{e,g}$, and integrating
over $y$ and $z$, we obtain
\begin{align}
\label{Eq:Full}
\notag
\Hc =&\sum_n  \left[\vphantom{\frac{1}{2}} -J (1+\eb (-1)^n)
  (\ah_{g,n}^\dagger\ah_{g,n+1} +\ah_{e,n}^\dagger\ah_{e,n+1} 
+ {\rm H.c.}) \right.
\\  &
 \left. + \frac{U_{gg}}{2} \ah_{g,n}^\dagger
 \ah_{g,n}^\dagger\ah_{g,n}\ah_{g,n} + 
 \frac{U_{ee}}{2} \ah_{e,n}^\dagger \ah_{e,n}^\dagger\ah_{e,n}\ah_{e,n}  \right.
\\ &
\notag
\left.  
+ 2 U_{ge}  \ah_{g,n}^\dagger\ah_{e,n}^\dagger\ah_{g,n}\ah_{e,n}
+ \frac{U_{ge}}{2} \ah_{g,n}^\dagger\ah_{g,n}^\dagger\ah_{e,n}\ah_{e,n}
\right.
\\ &
\notag
\left.
+ \frac{U_{ge}}{2}
\ah_{e,n}^\dagger\ah_{e,n}^\dagger\ah_{g,n}\ah_{g,n}
 -\tilde{\Delta} \ah_{g,n}^\dagger\ah_{g,n}  \right.
\\ &
\notag
\left.
-\mu(\ah_{g,n}^\dagger\ah_{g,n} +
\ah_{e,n}^\dagger\ah_{e,n})
\vphantom{\frac{1}{2}}\right]
\end{align}
where $\tilde{\Delta}>0$ is the energy difference between the ground and
excited orbitals (in the limit of tight confinement, $\tilde{\Delta}=\hbar
\omega_y$), and we have shifted the chemical potential. 
The nearest-neighbor hopping expressed as
-$\int dx w^*_n(x) \left( \frac{-\hbar^2}{2m}\frac{d^2}{dx^2}+V_{\rm
    lat}(x)\right)
w_{n+1}(x)$
is $J(1+\bar{\epsilon})$ for $n$ even and $J(1-\bar{\epsilon})$
for $n$ odd (hopping further than nearest-neighbors is dropped).
We also define
$U_{\alpha \beta} = g \int d^3 r \; \phi^2_\alpha(y,z)
\phi^2_\beta(y,z) |w_n(x)|^4$.  A similar analysis to the above is
carried out in \cite{zhou11}.

The Gross-Pitaevskii equation corresponding to (\ref{Eq:Full})
will have the solution $a_{e,n} = \sqrt{(\mu+2t)/U_{ee}}\equiv
\bar{a}_e$, $a_{g,n}=0$ which corresponds to all bosons being in the
excited band.  Inserting  $\ah_{e,n} = \bar{a}_e +\delta \ah_{e,n} $
into  (\ref{Eq:Full}) and expanding to quadratic order in $\delta \ah_{e,n} $
and $\ah_{g,n} $, we find (dropping the constant term)
\begin{align}
\label{Eq:HB}
\Hc_{\rm B}= \Hc_g + \Hc_e
\end{align}
where 
\begin{align}
\label{Eq:Hg}
\Hc_g &=\sum_n  \left[\vphantom{\frac{1}{2}} -J (1+\eb (-1)^n)
  ( \ah_{g,n}^\dagger \ah_{g,n+1} + {\rm H.c.}) \right. 
\\ & \left.
+(2J+ U_{ge} n_0-\Delta) \ah_{g,n}^\dagger  \ah_{g,n}  
+ \frac{U_{ge}n_0}{2} (\ah_{g,n}  \ah_{g,n}  + {\rm
  H.c.}) 
\right]
\notag
\end{align}
and
\begin{align}
\label{Eq:He}
\Hc_e &=\sum_n  \left[\vphantom{\frac{1}{2}} -J (1+\eb (-1)^n)
  (\delta \ah_{e,n}^\dagger\delta \ah_{e,n+1} + {\rm H.c.}) \right. 
\\ & \left.
+(2J+ U_{ee} n_0) \delta \ah_{e,n}^\dagger\delta \ah_{e,n}  
+ \frac{U_{ee}n_0}{2} (\delta \ah_{e,n} \delta \ah_{e,n}  + {\rm H.c.}) 
\right].
\notag
\end{align}
In these equations, we have introduced $n_0\equiv |\bar{a}_e|^2$ and 
$\Delta\equiv \tilde{\Delta} + U_{ee} n_0 - U_{ge} n_0$.
Interestingly, at the quadratic level the dynamics of bosons in the
ground and excited band is decoupled.  It is straightforward 
to diagonalize  ${\Hc}_e$ and see that it is stable.  Furthermore,
retaining higher-energy bands will  yield additional stable and gapped modes
which are unimportant.  
The Hamiltonian ${\Hc}_g$ is analyzed in the manuscript where the
subscript $g$ is dropped and $U \equiv U_{ge}$.  

\section{Evolution from the excited band}

We now consider the evolution of the initial state where all atoms are
in the excited band:  $\langle \ah_{g,n}
\rangle=0$, $\langle \ah_{e,n} \rangle=\sqrt{n_0}$.  Since this
initial state is a solution of the Gross Pitaevskii equation, it will
be stationary at the classical level.  Quantum fluctuations,
which are contained in (\ref{Eq:HB}), will 
trigger the evolution.  The atom number per site as a function of
time is given by 
\begin{align}
\label{Eq:gain}
F_{n}(t) &= \bra{\psi(t)} (\ah^\dagger_{g,n} \ah_{g,n} +
\ah^\dagger_{g,e}\ah_{g,e})\ket{\psi(t)}  \\
&= n_0 + \bra{\psi(t)} \ah^\dagger_{g,n} \ah_{g,n} \ket{\psi(t)} 
+ \bra{\psi(t)} \delta \ah^\dagger_{e,n} \delta \ah_{e,n}
\ket{\psi(t)} 
\notag
\end{align}
where $\ket{\psi(t)}=e^{-\frac{i}{\hbar} \Hc_{\rm B} t} \ket{0}$ and $\ket{0}$ is
the vacuum state of $\ah_{g,n} $ and $\delta \ah_{e,n} $
bosons.  This expression is valid when
\begin{equation}
\label{Eq:ineq}
 \sum_n \bra{\psi(t)}( \ah^\dagger_{g,n} \ah_{g,n} + \delta
 \ah^\dagger_{e,n} \delta \ah_{e,n} ) \ket{\psi(t)} 
\ll n_0 N.
\end{equation}
If there is a dynamical instability, there will inevitably be a time at which
(\ref{Eq:ineq}) breaks down, but (\ref{Eq:gain}) will be valid before then.
In the manuscript, we investigate the behavior of 
$G_n(t)=\bra{\psi(t)} \ah^\dagger_{g,n} \ah_{g,n} \ket{\psi(t)} $
which has exponential growth for a dynamical instability.

\bibliographystyle{apsrev4-1}
%

\end{document}